# Study of the $^{238}$U(d,p) surrogate reaction via the simultaneous measurement of gamma-decay and fission probabilities

Q. Ducasse[1,2], B. Jurado[1,*], M. Aïche[1], P. Marini[1], L. Mathieu[1], A. Görgen[3], M. Guttormsen[3], A. C. Larsen[3], T. Tornyi[3], J. N. Wilson[4], G. Barreau[1], G. Boutoux[5], S. Czajkowski[1], F. Giacoppo[3], F. Gunsing[6], T.W. Hagen[3], M. Lebois[4], J. Lei[7], V. Méot[5], B. Morillon[5], A. M. Moro[7], T. Renstrøm[3], O. Roig[5], S. J. Rose[3], O. Sérot[2], S. Siem[3], I. Tsekhanovich[1], G. M. Tveten[3], M. Wiedeking[8]

1) CENBG, CNRS/IN2P3-Université de Bordeaux, Chemin du Solarium B.P. 120, 33175 Gradignan, France
2) CEA-Cadarache, DEN/DER/SPRC/LEPh, 13108 Saint Paul lez Durance, France
3) Department of Physics, University of Oslo, 0316 Oslo, Norway
4) IPN d'Orsay, Bâtiment 100, 15 rue G. Clemenceau, 91406 Orsay Cedex, France
5) CEA DAM DIF, 91297 Arpajon, France
6) CEA Saclay, DSM/Irfu/SPhN, 91191 Gif-sur-Yvette Cedex, France
7) Departamento de FAMN, Universidad de Sevilla, Apartado 1065, 41080 Sevilla, Spain
8) iThemba LABS, P.O. Box 722, 7129 Somerset West, South Africa

**Abstract:** We investigated the $^{238}$U(d,p) reaction as a surrogate for the n + $^{238}$U reaction. For this purpose we measured for the first time the gamma-decay and fission probabilities of $^{239}$U* simultaneously and compared them to the corresponding neutron-induced data. We present the details of the procedure to infer the decay probabilities, as well as a thorough uncertainty analysis, including parameter correlations. Calculations based on the continuum-discretized coupled-channels method and the distorted-wave Born approximation (DWBA) were used to correct our data from detected protons originating from elastic and inelastic deuteron breakup. In the region where fission and gamma emission compete, the corrected fission probability is in agreement with neutron-induced data, whereas the gamma-decay probability is much higher than the neutron-induced data. We have performed calculations of the decay probabilities with the statistical model and of the average angular momentum populated in the $^{238}$U(d,p) reaction with the DWBA to interpret these results.

## I Introduction

Neutron-induced reaction cross sections of short-lived nuclei are important in several domains such as fundamental nuclear physics, nuclear astrophysics and applications in nuclear technology. These cross sections are key input information for modeling stellar element nucleosynthesis via the s and r-processes. They also play an essential role in the design of advanced nuclear reactors for the transmutation of nuclear waste, or reactors based on innovative fuel cycles like the Th/U cycle. However, very often these cross sections are extremely difficult (or even impossible) to measure due to the high radioactivity of the targets involved.

The surrogate-reaction method was first developed at the Los Alamos National Laboratory by Cramer and Britt [1]. This indirect technique aims to determine neutron-induced cross sections of reactions involving short-lived nuclei that proceed through the formation of a compound nucleus, i.e. a nucleus that is in a state of statistical equilibrium. In this method, the same compound nucleus as in the neutron-induced reaction of interest is produced via an alternative, or surrogate, reaction (e.g. a transfer or inelastic scattering reaction). The surrogate-reaction method is schematically represented in Fig. 1. The left part of Fig. 1 illustrates a neutron-induced reaction on target $A$-1, which leads to the formation of nucleus $A^*$ at an excitation energy $E^*$. The nucleus $A^*$ can decay via different exit channels: fission, gamma-decay, neutron emission, etc. On the right part of Fig. 1, the same compound nucleus $A^*$ is produced via a surrogate reaction. In Fig. 1, the surrogate reaction is a transfer reaction between a projectile $y$ (a light nucleus) and a target $X$, leading to the heavy recoil nucleus $A^*$ and an ejectile $w$. The charge and mass identification of the ejectile $w$ allows one to deduce the charge and mass of the decaying nucleus $A^*$, and the measurement of the ejectile kinetic energy and emission angle provides its excitation energy $E^*$. In most applications of the surrogate method, the surrogate reaction is used to measure the decay probability $P_\chi$ and the desired neutron-induced reaction cross section is "simulated" by applying the equation:

$$\sigma_\chi^{A-1}(E_n) = \sigma_{CN}^A(E_n) \cdot P_\chi^A(E^*) \tag{1}$$

where the index $\chi$ represents the decay mode (e.g. fission or gamma-ray emission) and $\sigma_{CN}^A(E_n)$ is the cross section for the formation of a compound-nucleus $A^*$ by the absorption of a neutron of energy $E_n$ by nucleus $A$-1. The compound-nucleus formation cross section $\sigma_{CN}^A(E_n)$ can be calculated with phenomenological optical-model calculations with an accuracy of about 10% for nuclei not too far from the stability valley [2]. The excitation energy $E^*$ and the neutron energy $E_n$ are related via the equation $E^* = S_n + \frac{A-1}{A}E_n$, where $S_n$ is the neutron separation energy of nucleus $A$.

---

[*] jurado@cenbg.in2p3.fr

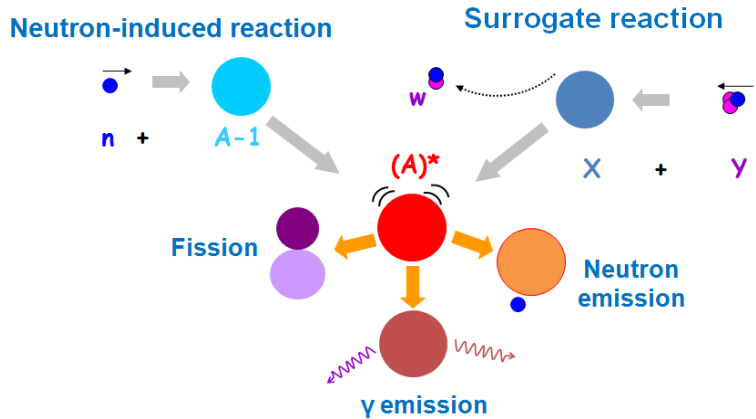

**Figure 1:** (Color Online) Schematic representation of the surrogate-reaction method. The surrogate reaction depicted here is a transfer reaction $X(y,w)A^*$. Three of the possible exit channels (fission, gamma emission and neutron emission) are represented.

One of the main advantages of the surrogate-reaction method is that, in some cases, one can find a surrogate reaction where the target $X$ is stable or less radioactive than the target $A$-1. However, the interest of the surrogate-reaction method goes well beyond the accessibility of the targets in direct-kinematics experiments. Indeed, due to the current impossibility to produce free-neutron targets, surrogate reactions might be used to simulate neutron-induced reactions of very short-lived nuclei that are only available as radioactive beams. Of particular interest is the (d,p) reaction, i.e. the transfer of a neutron from the weakly bound deuteron target to the radioactive beam, which intuitively appears as the closest reaction to a neutron-induced reaction in inverse kinematics.

For the surrogate-reaction method to be valid, several conditions have to be fulfilled [2]. First, both the neutron-induced and the surrogate reactions must lead to the formation of a compound nucleus. In that case the decay of nucleus $A^*$ is independent of the entrance channel and the reaction cross section can be factorized into the product of the compound-nucleus formation cross section and the decay probability into a channel $\chi$, as in eq. (1). The second condition is that the decay probability measured in the surrogate reaction has to be equal to the decay probability in the neutron-induced reaction. This is the case in at least two limiting situations: if the angular momentum ($J$) and parity ($\pi$) distributions populated in the neutron- and transfer-induced reactions are the same, or if the decay probability of the compound nucleus is independent of its angular momentum and parity, which is the so-called Weisskopf-Ewing limit. Since for most surrogate reactions it is not yet possible to determine the populated $J^\pi$ distribution [2], the validity of the surrogate method has to be verified "a posteriori", by comparing the obtained results with well known neutron-induced data.

Surrogate-reaction studies performed in the last decade have shown that fission cross sections obtained via the surrogate-reaction method are generally in good agreement with the corresponding neutron induced data, see e.g. [3] and other examples included in [2]. However, discrepancies as large as a factor 10 have been observed when comparing radiative-capture cross sections of rare-earth nuclei obtained in surrogate and neutron-induced reactions [4, 5]. These significant differences have been attributed to the higher angular momenta populated in the surrogate reaction. At excitation energies close to $S_n$, neutron emission is very sensitive to the angular momentum of the decaying nucleus $A^*$, as only the ground state and the first excited states of the residue nucleus $A$-1 can be populated. When the angular momentum of $A^*$ is considerably higher than the angular momentum of the first states of $A$-1, neutron emission is hindered and the nucleus $A^*$ predominantly decays by gamma emission, which is the only open decay channel [5]. This effect is expected to be reduced for actinides, as they have more low-lying states than rare-earth nuclei, thus making neutron decay less selective. However, the radiative-capture cross section of $^{232}$Th obtained via the $^{232}$Th(d,p) surrogate reaction in ref. [6] shows very large discrepancies with respect to the neutron-induced radiative-capture cross section at low neutron energies.

Similarly to the situation at energies close to the ground state, the energy region close to the fission barrier is also characterized by a low density of states and a significant dependence of the fission probability on the angular momentum is expected by theory [2]. Therefore, it is surprising that the spin/parity mismatch between the surrogate and neutron-induced reactions has no major impact on the measured fission probabilities. To shed light into this puzzling observation, it is first of all necessary to demonstrate the much weaker sensibility of the fission probability to angular momentum by simultaneously measuring fission and gamma-decay probabilities for the same nucleus at the same excitation energy. This has never been done before and is the aim of the present work. Here we concentrate on the $^{238}$U(d,p) reaction, which is used to simulate the n+$^{238}$U reaction for which good-quality neutron-induced data on fission and capture cross sections exist. The measurement of the gamma-decay probability at excitation energies where the fission channel is open is challenging because of the background of gamma rays emitted by the fission fragments. Above $S_n$, the gamma-decay probability of $^{239}$U decreases very rapidly with excitation energy, whereas the fission probability increases. Therefore, the fraction of gamma rays coming from the fission fragments increases gradually with $E^*$ until they represent most of the detected gamma rays.

In this work, we restricted the measurement of the gamma-decay probability to the range $E^* < S_n$ +1.5 MeV, in order to limit the uncertainty due to the subtraction of the fission-fragment gamma-ray background.

The (d,p) reaction presents a difficulty. Britt and Cramer [7] noticed that, above a certain excitation energy, the fission cross sections obtained via the (d,p) surrogate reaction were significantly lower than the corresponding neutron-induced cross section. They attributed this to the elastic breakup of the deuteron. Deuteron breakup is actually a rather complex process and has recently been the subject of several theoretical works, see e.g. [8, 9]. In the present work, we use the approach of [9] to correct our data from the effects of deuteron breakup.

Contrary to the internal surrogate-ratio method used by Allmond et al. [10], the technique employed in the present work for the extraction of the gamma-decay probability of fissile nuclei does not require the knowledge of the fission cross section and of the complete level scheme of nucleus *A*. Our method is of more general interest than the one of Allmond et al., as it can be applied to short-lived fissile nuclei for which no experimental information is available.

## II Experiment

The experiment was performed at the Oslo Cyclotron Laboratory that provided a deuteron beam of 15 MeV energy with an intensity of about 4 enA. The setup is sketched in Fig. 2. The multi-strip ΔE/E silicon telescope SiRi [11] was used to identify the ejectiles and determine their kinetic energies and angles. SiRi covered polar angles ranging from 126 to 140 in steps of 2 degrees. An ensemble of four PPACs [12], located at forward angles, was used to detect the fission fragments in coincidence with the ejectiles. The reaction chamber housing SiRi, the PPACs and the $^{238}$U target was surrounded by the CACTUS array [13], constituted of 27 high-efficiency NaI detectors placed 22 cm away from the target. CACTUS was used to detect gamma rays with energies ranging from few hundreds of keV to about 10 MeV emitted in coincidence with the ejectiles. A 21 µm thick aluminum foil was placed in front of the SiRi telescope to stop fission fragments.

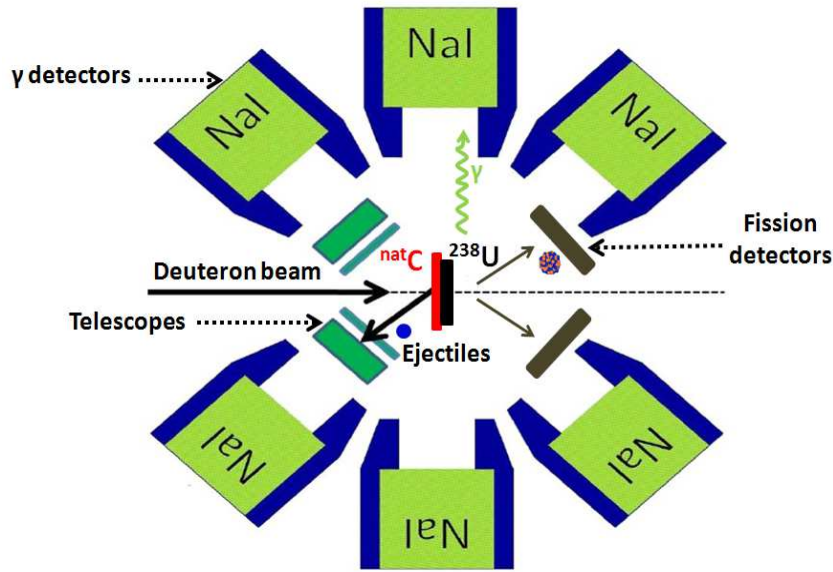

**Figure 2:** (Color Online) Schematic view of the setup used at the Oslo Cyclotron Laboratory for the simultaneous measurement of fission and gamma decay-probabilities.

The amplified signals of the telescopes, and the fission and gamma detectors were digitized with an analog-to-digital converter. All the detector signals were pulse-shaped into fast timing signals and sent to a time-to-digital converter to measure the time differences between the telescopes and the fission and gamma detectors. The acquisition system was triggered by a logic OR of the $\Delta E$-$E$ coincidences of each telescope strip. We used a high-quality metallic $^{238}$U target, with 99.5% isotopic purity, produced by the GSI target laboratory. It had a thickness of 260 µg/cm$^2$ and was deposited on a 40 µg/cm$^2$ natural carbon layer. Great care was taken to avoid as much as possible the oxidation of the target, which was produced only a few days before the measurement and was transported from GSI to Oslo under vacuum conditions.

## III Data analysis

The decay probability in the outgoing channel $\chi$ of the $^{239}$U* nucleus produced in the $^{238}$U(d,p) reaction can be obtained as :

$$P_\chi(E^*) = \frac{N_\chi^C(E^*)}{N^S(E^*) \cdot \varepsilon_\chi(E^*)} \qquad (2)$$

Here $N^S(E^*)$ is the so-called "singles spectrum", i.e. the total number of detected protons as a function of excitation energy $E^*$. $N_\chi^C(E^*)$ is the "coincidence spectrum", corresponding to the number of protons detected in coincidence with the observable that identifies the decay mode, e.g. a fission fragment or a gamma ray, and $\varepsilon_\chi$ is the associated detection efficiency. In the absence of protons originating from contaminant reactions, the quantity $N^S(E^*)$ corresponds to the total number of formed $^{239}$U* nuclei and $N_\chi^C(E^*)/\varepsilon_\chi$ to the number of $^{239}$U* nuclei that have decayed via channel $\chi$. The following sections discuss how the quantities involved in eq. (2) are obtained.

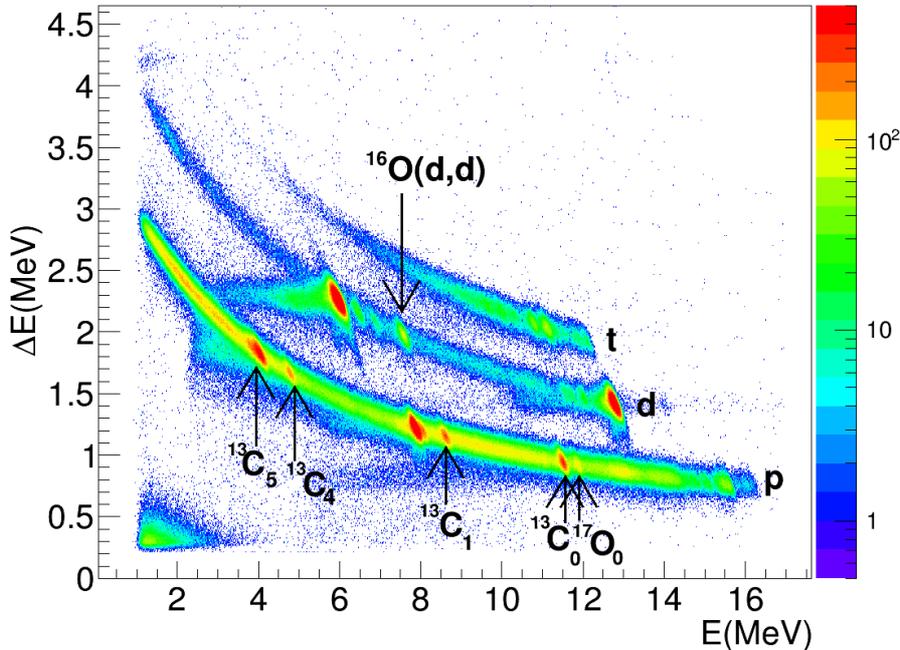

**Figure 3:** (Color Online) Energy loss versus residual energy of the ejectile measured at 126 degrees. The ejectiles corresponding to different hydrogen isotopes are indicated. The arrows in the lower part indicate the $^{17}$O and $^{13}$C states used for the energy calibration of the telescopes, where $^{13}$C$_0$ corresponds to the ground state of $^{13}$C, $^{13}$C$_1$ to the first excited state, etc.

## A Excitation energy

The excitation energy of $^{239}$U* is determined from the measured kinetic energy and emission angle of the protons, by applying energy and momentum conservation laws. Figure 3 shows an identification plot representing the energy loss in the ΔE detector as a function of the residual energy in the E detector of the ejectiles measured in one strip of the silicon telescope. The different ejectiles corresponding to different transfer channels (and different uranium isotopes) can be well distinguished. Interactions of the deuteron beam with oxygen contamination and the carbon backing of the target lead to the production of O and C isotopes in their ground and excited states. Some of those states can be observed as well-separated peaks in the identification plot, and correspond to well-defined energies of the ejectiles. The ejectile energies corresponding to the formation of $^{17}$O in the ground state by the $^{16}$O(d,p) reaction and several $^{13}$C states populated in the $^{12}$C(d,p) reaction have been used to calibrate in energy the SiRi telescope. The calibration procedure was validated by comparing our calibrated singles spectrum with the spectrum measured by Erskine [14]. The excitation-energy resolution was estimated from the standard deviation of the peaks associated to the $^{238}$U(d,d') reaction and amounts to about 50 keV.

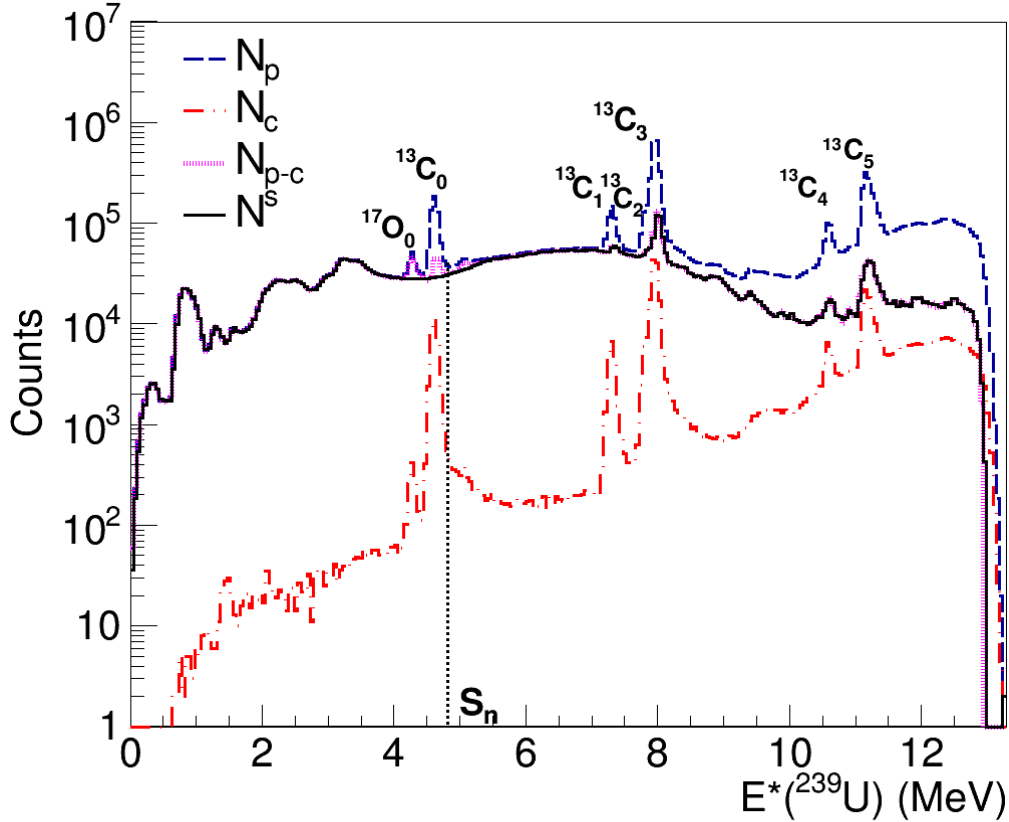

**Figure 4:** (Color Online) Number of detected protons as a function of the excitation energy of $^{239}$U* measured at 126 degrees. The blue dashed line is the proton spectrum, $N_p$. The red dashed-dotted line, $N_c$, corresponds to the spectrum obtained with the carbon backing without normalization factor. The spectrum obtained after subtraction of the carbon spectrum, $N_{p-c}$, is represented by the pink dotted line. The singles spectrum $N^S$ is shown as a black solid line. The vertical dotted line represents the neutron separation energy of $^{239}$U. The peaks related to the formation of $^{17}$O and $^{13}$C in the ground- and first excited states are indicated with the same notation as in Fig. 3.

**B Singles spectrum**

To determine the singles spectrum we first selected the protons via the identification plot shown in Fig. 3 and represented the number of protons as a function of the excitation energy of $^{239}$U*. This spectrum is called the proton spectrum $N_p$ and is represented in blue in Fig. 4. The peaks above 4 MeV correspond to protons originating from reactions on the carbon backing and the oxygen of the target. The different steps undertaken to remove these contaminant events are illustrated in Fig. 4. First we subtracted from the $N_p$ spectrum the proton spectrum, appropriately normalized, obtained in a separated measurement with the carbon backing only, the $N_c$ spectrum. The spectrum that results from the subtraction is labelled as $N_{p\text{-}c}$ in Fig. 4. Because the shape of the carbon peaks in the $N_p$ and $N_c$ spectra was not identical, the carbon peaks could not be completely removed from $N_p$, as can be seen in Fig. 4. To remove these peak residues and the oxygen peaks, for which a background measurement cannot be performed, the $N_{p\text{-}c}$ spectrum was interpolated below the contaminant peaks with a polynomial function. To determine the shape of the polynomial we exploited the angular dependence of the kinetic energy of the emitted protons. This dependence is much stronger for protons ejected in reactions on light nuclei such as C and O than on the heavy $^{238}$U nucleus. Therefore, the contaminant peaks move to higher excitation energies of $^{239}$U* as the detection angle increases. Thus, the shape of the singles spectrum below the contaminant peaks at a given angle was deduced from the shape of the proton spectrum measured at a different angle. The interpolation procedure was only applied in the vicinity of $S_n$, which is the region of interest in the present work. The singles spectrum $N^s$ is shown as the solid black line in Fig. 4, it represents the spectrum of protons coming from reactions on the $^{238}$U target.

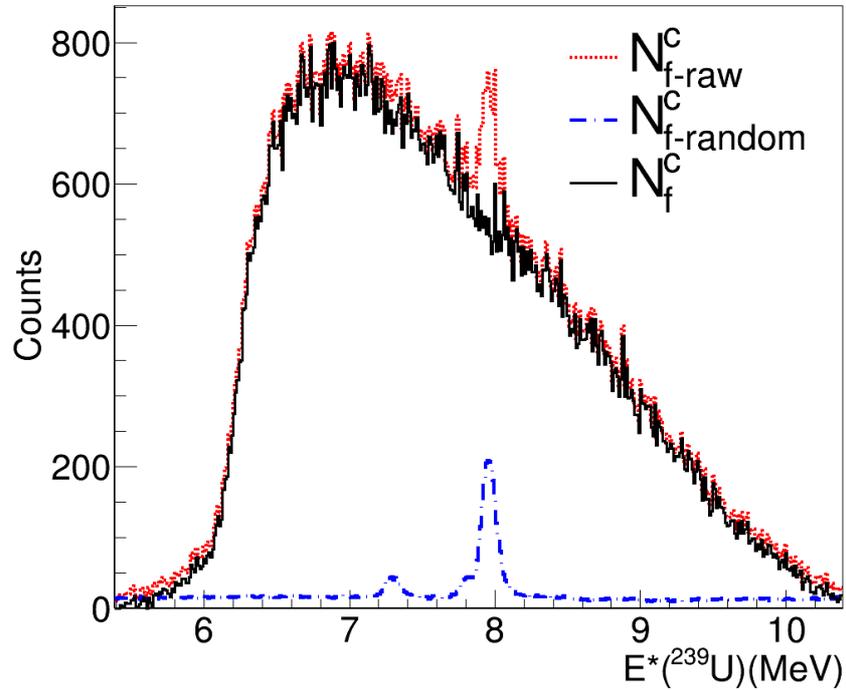

**Figure 5:** (Color Online) The red dotted line, $N^c_{f-raw}$, is the spectrum of protons detected in coincidence with a signal in the fission detector as a function of the excitation energy of $^{239}$U*. The blue dashed-dotted line corresponds to the normalized random-coincidence spectrum, $N^c_{f-random}$. The fission coincidence spectrum, $N^C_f$, is shown as the solid black line.

**C Fission coincidence spectrum**

Figure 5 shows the spectrum $N^c_{f-raw}$ that results from selecting the protons detected in coincidence with a signal in any of the four PPACs. This spectrum presents an intense peak at about 8 MeV excitation energy. This peak corresponds to random coincidences between protons originating from reactions on the carbon backing and the fission detectors. To remove these random events we subtracted from the coincidence spectrum the spectrum (properly normalized) obtained when selecting the events with a time difference lying outside of the coincidence window, $N^c_{f-random}$, see [15] for details. The result is the fission coincidence spectrum $N^C_f(E^*)$, shown as a solid black line in Fig. 5.

**D Fission-detection efficiency**

The fission-detection efficiency, $\varepsilon_f(E^*)$, is the last term of eq. (2) needed to infer the fission-decay probability of $^{239}$U*. The fission-detection efficiency is determined by the solid angle covered by the four PPACs and by the angular anisotropy of the fission fragments in the laboratory reference system. The latter is given by the angular anisotropy of the fragments in the center of mass (CM) system corrected for kinematical effects due to the recoil energy of the fissioning nucleus.

The solid angle was measured with a $^{252}$Cf source of known activity and was found to be (41.1 ± 0.3)% of 2π. The PPACs used in this experiment were not position-sensitive detectors and the angular anisotropy in the CM could not be measured. Therefore, we used the angular anisotropy in the CM measured by Britt and Cramer [7] for the $^{238}$U(d,p) reaction at 18 MeV deuteron incident energy. To include the angular anisotropy effects, we performed a Monte-Carlo simulation that reproduces the geometrical efficiency. In the simulation, the velocities of the fission fragments in the CM were taken from the GEF code [16]. The total efficiency obtained with our simulation is $\varepsilon_f$ = (48.0 ± 3.5)%. We considered a constant efficiency since the variation of the efficiency with the excitation energy is very weak and is largely included in the final uncertainty. The final uncertainty on the fission efficiency is dominated by the uncertainty on the angular anisotropy in the CM that has been numerically propagated into the final efficiency via the Monte-Carlo simulation.

**E Gamma-coincidence spectrum**

To obtain the gamma-decay probability we need to determine the number of formed $^{239}$U* nuclei that decay through a gamma-ray cascade, $N^C_\gamma(E^*)$, i.e. the number of $^{239}$U* nuclei that de-excite by emitting gamma rays only. The efficiency of the CACTUS array is about 14.5% at 1.33 MeV gamma-ray energy. Therefore, in most cases, we detected only one gamma ray per cascade. For the few cases where more than one NaI detector was hit in one event, we randomly selected one detector signal amplitude in the offline data analysis. In that way, we avoided counting more than one gamma ray per cascade.

To calibrate in energy the NaI scintillators we used the gamma rays emitted in the de-excitation of several excited states of $^{13}$C and $^{17}$O populated by the (d,p) reaction. As mentioned in the introduction, we restricted the measurement of the gamma-decay probability to $E^* < S_n +1.5$ MeV. For this reason, a threshold, $E^{th}_\gamma$ =1.5 MeV, was applied to the detected gamma rays in order to eliminate the gamma rays originating from the residue nucleus $^{238}$U* produced after neutron emission from $^{239}$U*. This threshold is shown in the two-dimensional plot in Fig. 6 representing the excitation energy of $^{239}$U* versus the measured gamma-ray energy. The gamma rays emitted by the $^{238}$U* residue are on the left side of the diagonal line which intersects the $E^*$ axis at $S_n$, whereas the gamma rays emitted by the $^{239}$U* nucleus are on the left side of the diagonal line with origin at $E^*=0$. The region used for the determination of the gamma-decay probability is represented by the red dashed line. The spectrum $N_\gamma^{C,tot}$ corresponding to the coincidences between protons and one detected gamma ray with energy above $E_\gamma^{th}$ is shown as a blue dashed-dotted line in Fig. 7. The same procedure as the one described in Section C was used to remove random coincidences. As expected, the coincidence spectrum $N_\gamma^{C,tot}$ shows a step decrease at $S_n$ because neutron emission starts to compete with gamma emission.

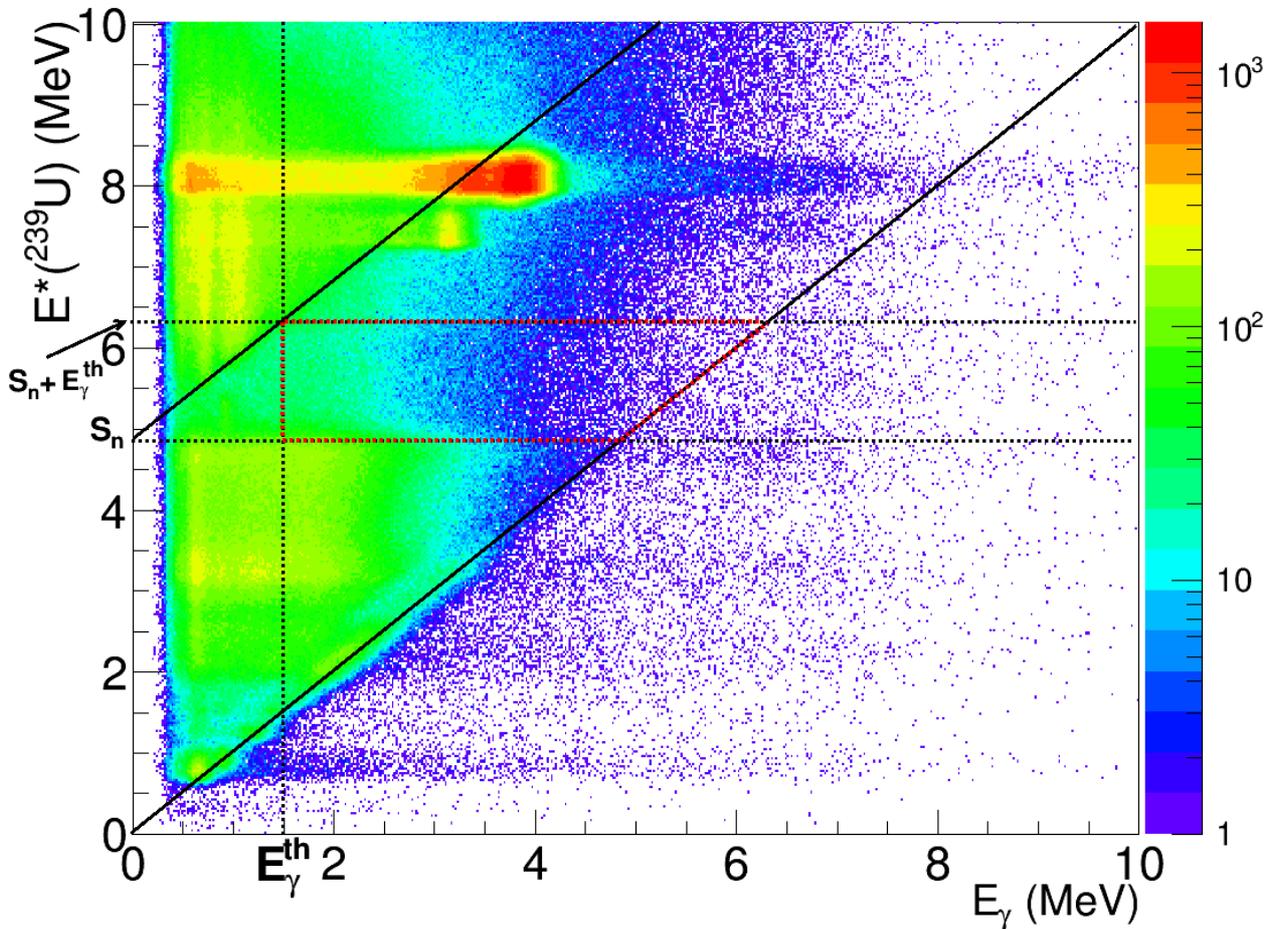

**Figure 6:** (Color online) Excitation energy of $^{239}$U versus detected gamma-ray energy. The applied energy threshold, $E_\gamma^{th}$, is represented by the vertical dotted line. Excitation energies corresponding to $S_n$ and $S_n + E_\gamma^{th}$ are indicated by horizontal dotted lines. The 45-degree lines with origin at $E^* = 0$ and $E^* = S_n$ are represented by full lines. The region used in the analysis of the gamma-decay probability is highlighted by the red dashed line.

A time window of 11 ns was used for selecting the proton-gamma-ray coincidences. This time window in combination with the 1.5 MeV gamma-ray energy threshold allowed us to remove the majority of the contaminant gamma rays emitted by the Na and I nuclei of the scintillator material after the capture of a neutron emitted by $^{239}$U*. Indeed, in the excitation-energy range of interest, the maximum kinetic energy carried by the neutron is $E_{n,max} = E^* - S_n = 1.5$ MeV. Consequently, only neutrons with lower kinetic energies can be captured in the NaI detectors. Taking into account the time resolution of the CACTUS NaI detectors of about 10 ns and the average interaction distance of the neutrons in the NaI crystals of 25 cm [6], the time window of 11 ns suppresses 95% of the emitted neutrons with $E_n \leq 0,360$ MeV and 68% with $E_n \leq 1$ MeV. Above a few hundred keV the neutron inelastic cross sections of Na and I are one or more orders of magnitude larger than the capture cross sections, but the gamma rays originating from inelastic scattering on Na and I are also removed by the 1.5 MeV gamma-ray energy threshold. To demonstrate the absence of gamma rays coming from the interaction of neutrons emitted by $^{239}$U* with the NaI detectors, we analyzed the data using a time window of 24 ns. Using this window would in principle lead to an increase of the gamma-decay probability due to the presence of more contaminant gamma rays coming from neutron capture in the NaI. However, the results agree within the error bars. In fact, the contribution from capture events in the NaI starts to be significant only when a time window as large as 42 ns is used. The latter window includes a large contribution of neutrons with $E_n < 200$ keV for which the capture cross sections are rather high.

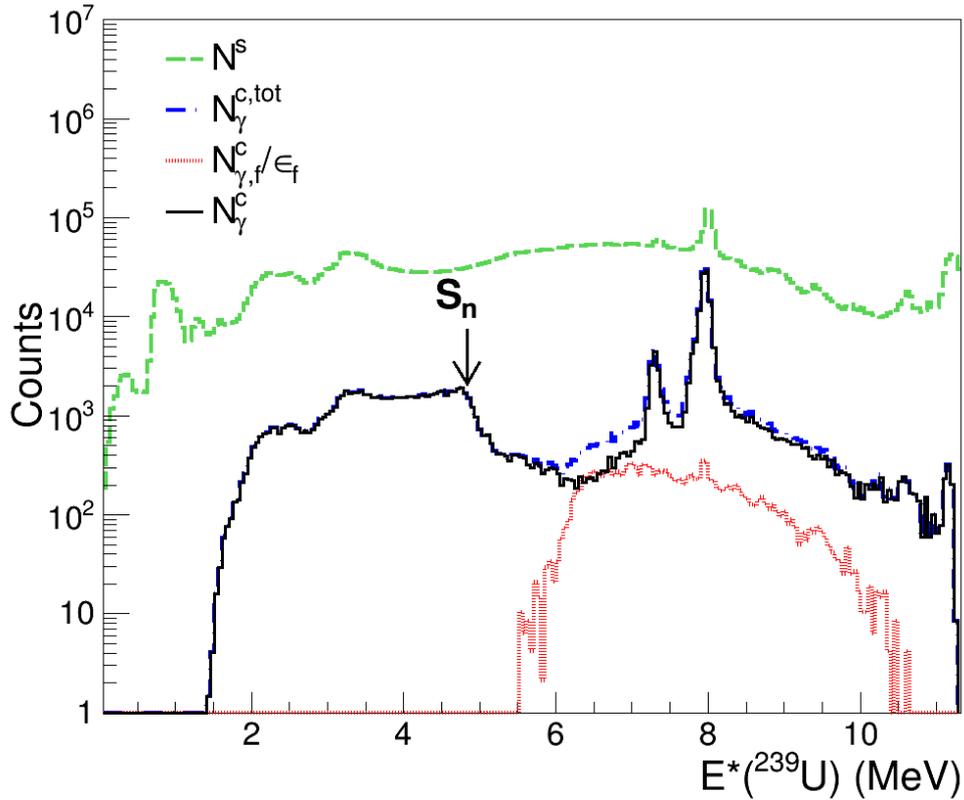

**Figure 7:** (Color online) The blue dashed-dotted line $N_\gamma^{C,tot}$ represents the number of protons detected in coincidence with a gamma ray detected in any of the CACTUS detectors as a function of the excitation energy of $^{239}$U measured at 126 degrees. A threshold in the gamma-ray energy $E_\gamma^{th}$ > 1.5 MeV and a time window of 11 ns were used to obtain this spectrum. The red-dotted line is the fission-gamma coincidence spectrum $N_{\gamma,f}^C$ divided by the fission efficiency $\varepsilon_f$. The black line represents the gamma-coincidence spectrum $N_\gamma^C$. The green dashed line is the singles spectrum $N^S$. The arrow indicates the neutron separation energy of $^{239}$U*.

The total gamma-coincidence spectrum $N_\gamma^{C,tot}$ has to be corrected for the prompt gamma rays emitted by the fission fragments. The impact of gammas originating from the fission fragments can be noticed on Fig. 7, where we observe an increase of the coincidence spectrum, $N_\gamma^{C,tot}$, above about 6 MeV close to the onset of fission. This correction can be done by measuring triple coincidences between protons, fission fragments and gamma rays. The corrected coincidence spectrum $N_\gamma^C$ is then obtained as:

$$N_\gamma^C(E^*) = N_\gamma^{C,tot}(E^*) - \frac{N_{\gamma,f}^C(E^*)}{\varepsilon_f(E^*)} \quad (3)$$

where $N_{\gamma,f}^C$ is the number of gamma cascades detected in coincidence with a proton and a fission fragment, and $N_\gamma^C$ is the final gamma-coincidence spectrum shown as a solid line in Fig. 7.

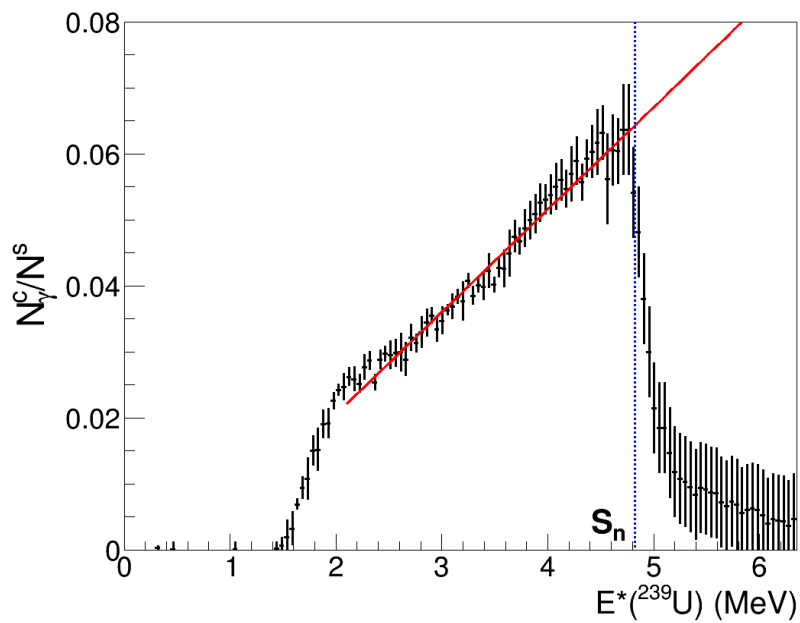

**Figure 8:** (Color online) Ratio between the gamma-coincidence and the singles spectra. The vertical dotted line indicates the neutron separation energy of $^{239}$U and the red solid line is a linear fit to the data in the $E^*$ interval [2 MeV; $S_n$].

**F Gamma-cascade detection efficiency**

To obtain the gamma-decay probability, one needs to determine the efficiency for detecting a gamma cascade rather than the efficiency for detecting a gamma of a particular energy. In this work, we used the EXtrapolated Efficiency Method (EXEM) developed in [17] to determine the gamma-cascade detection efficiency. In a surrogate reaction it is possible to populate excitation energies below the neutron separation energy. For a neutron-rich nucleus as $^{239}$U*, that does not fission nor emits protons below $S_n$, the only possible de-excitation mode at $E^* < S_n$ is gamma decay. Therefore, the gamma-decay probability is equal to 1:

$$P_\gamma(E^*) = 1 = \frac{N_\gamma^C(E^*)}{N^S(E^*) \cdot \varepsilon_\gamma(E^*)} \quad \text{for } E^* < S_n \quad (4)$$

From eq. (4) it follows that:

$$\varepsilon_\gamma(E^*) = \frac{N_\gamma^C(E^*)}{N^S(E^*)} \quad \text{for } E^* < S_n \tag{5}$$

Thus, for excitation energies below $S_n$, the gamma-cascade detection efficiency, $\varepsilon_\gamma(E^*)$, can be directly obtained from the ratio between the gamma-coincidence and the singles spectra. For medium-mass and actinide nuclei in the region of continuum level densities there is no reason to expect a drastic change at $S_n$ of the characteristics of the gamma cascades (multiplicity and average gamma energy), and thus of $\varepsilon_\gamma(E^*)$. This is the main idea on which the EXEM is based. The EXEM assumes that the dependence of the gamma-cascade detection efficiency $\varepsilon_\gamma$ on $E^*$ measured below $S_n$ can be extrapolated to excitation energies above $S_n$. This is illustrated in Fig. 8, where the ratio of the gamma-coincidence and the singles spectra is shown together with a linear fit. The values of the fit function evaluated at $E^*$ above $S_n$ gave us the gamma-cascade efficiency used to determine the gamma-decay probability. In the excitation-energy range of interest, the gamma-cascade detection efficiency increases from a value of about (6.5±0.5)%, near $S_n$, to about (8.5±0.7)% at $E^*$=6.3 MeV. The uncertainty on $\varepsilon_\gamma$ above $S_n$ was obtained from the uncertainties on the fit parameters.

The validity of the EXEM applied to the actinide region is demonstrated in [18] where we present statistical-model calculations performed with the EVITA code (see section IV B) of the average gamma energy and multiplicity as a function of excitation energy for $^{239}$U*. These calculations agree rather well with the experimental values below $S_n$. Above $S_n$, the calculations show that there is no change in the slope of these two quantities and that the linear increase of the efficiency is mainly due to a linear increase of the average multiplicity. In addition, in [18] we further demonstrate the validity of the EXEM with the study of the $^{239}$Np fissile nucleus produced in the $^{238}$U($^3$He,d) reaction.

**G Uncertainty analysis**

Considering eq. (2), the relative uncertainty of $P_\chi$ at a given $E^*$ is given by:

$$\frac{Var(P_\chi(E^*))}{(P_\chi(E^*))^2} = \frac{Var(N_\chi^C(E^*))}{(N_\chi^C(E^*))^2} + \frac{Var(N^S(E^*))}{(N^S(E^*))^2} + \frac{Var(\varepsilon_\chi(E^*))}{(\varepsilon_\chi(E^*))^2}$$
$$-2\cdot\frac{Cov(N^S(E^*);N_\chi^C(E^*))}{N_\chi^C(E^*)\cdot N^S(E^*)} - 2\cdot\frac{Cov(N_\chi^C(E^*);\varepsilon_\chi(E^*))}{N_\chi^C(E^*)\cdot\varepsilon_\chi(E^*)} + 2\cdot\frac{Cov(N^S(E^*);\varepsilon_\chi(E^*))}{N^S(E^*)\cdot\varepsilon_\chi(E^*)} \tag{6}$$

where *Var* and *Cov* represent the variance and the covariance of the measured quantities, respectively. We have shown in [15] that for the fission probability our experimental procedure allows us to disregard the two last covariance terms in eq. (6), but that the covariance between single and coincidence events has a significant impact on the final uncertainty. In this work, we present a procedure to determine the covariance terms between the measured quantities that is different from the mathematical procedure described in [15]. Our new procedure allowed us to extract the covariance terms associated to the gamma-decay probability in a straightforward manner. For simplicity, in the following equations we will omit the dependence on $E^*$ of all the measured quantities.

To illustrate our alternative approach we will consider the case of $Cov(N^S; N_f^C)$, keeping in mind that this procedure can be used to obtain the covariance of any other two quantities involved in the measurement of the decay probabilities. The covariance $Cov(N^S; N_f^C)$ is a measure of how fluctuations in $N^S$ affect the value of $N_f^C$. One way to determine it is by repeating the measurements in exactly the same experimental conditions (geometry, beam intensity, measuring time, etc.) and by representing the measured $N^S$ versus $N_f^C$. Even though the experimental conditions are identical, $N^S$ and $N_f^C$ will fluctuate, because they are random variables that follow Poisson statistics. Of course, this procedure is generally not done. Alternatively, one can use the data collected during the experiment to construct groups of independent measured events with values for $N^S$ that are sampled from a Gaussian distribution centered at a given value of $<N^S>$ (e.g. 200) and with a standard deviation equal to $\sqrt{<N^S>}$. In this way, one "simulates" how $N^S$ would have varied if one would have performed exactly the same experiment many times. Figure 9 a) shows the impact of the variation of $N^S$ on the measured values of $N_f^C$ and Fig. 9 b) shows the impact of varying the quantity $N^{AC} = N^S - N_f^C$ on the measured values of $N_f^C$. The quantities $N^{AC}$ and $N_f^C$ are uncorrelated and their corresponding covariance term is zero. The comparison of parts a) and b) of Fig. 9 allows one to asses the differences in the characteristic pattern of two variables that are correlated (Fig. 9 a) and uncorrelated (Fig. 9 b). To obtain the plots of Fig. 9, we have used independent groups of experimental data in an $E^*$ region free of events coming from contaminant reactions. The variance and covariance terms are then determined with the estimators:

$$Var(N_f^C) = \frac{1}{n}\sum_{i=1}^{n}\left(N_{fi}^C - \langle N_f^C \rangle\right)$$

$$Cov(N^S; N_f^C) = \frac{1}{n}\sum_{i=1}^{n}\left(N_i^S - \langle N^S \rangle\right)\left(N_{fi}^C - \langle N_f^C \rangle\right)$$

(7)

where $n$ is the number of groups of data (or the number of points on Fig. 9) and the average quantities $<N>$ are given by $\langle N \rangle = \frac{1}{n}\sum_{i=1}^{n} N_i$. When we apply this experimental procedure we obtain $Cov(N^S; N_f^C) \approx Var(N_f^C)$, in agreement with the result obtained in [15].

For the gamma-decay probability, the two last covariance terms in eq. (6) cannot be in principle neglected because $\varepsilon_\gamma$ has been obtained with the EXEM which involves the $N^S$ and $N_\gamma^C$ variables. Moreover, because of the subtraction of prompt-fission gamma rays, we have to consider three additional covariance terms. Indeed, from eq. (3) it follows:

$$Var(N_\gamma^C) = Var(N_\gamma^{C,tot}) + \frac{Var(N_{\gamma,f}^C)}{\left(\varepsilon_f\right)^2} + \left(\frac{N_{\gamma,f}^C}{\varepsilon_f}\right)^2 \cdot \frac{Var(\varepsilon_f)}{\varepsilon_f^2} - 2\cdot\frac{Cov(N_\gamma^{C,tot}; N_{\gamma,f}^C)}{\varepsilon_f}$$
$$+2N_{\gamma,f}^C \cdot \frac{Cov(N_\gamma^{C,tot}; \varepsilon_f)}{\varepsilon_f^2} - 2N_{\gamma,f}^C \cdot \frac{Cov(N_{\gamma,f}^C; \varepsilon_f)}{\varepsilon_f}$$

(8)

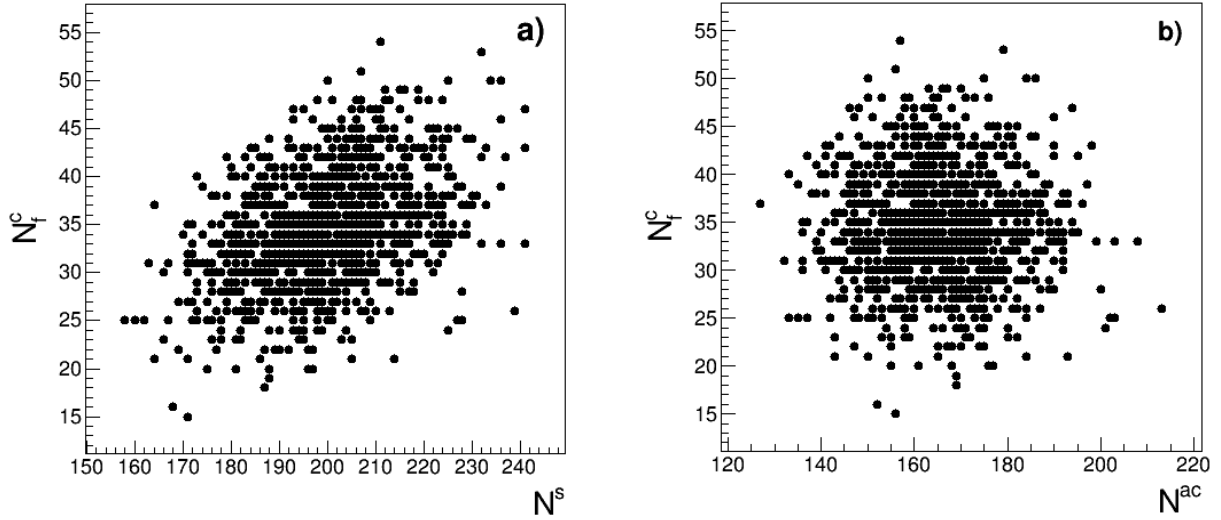

**Figure 9:** Measured $N_f^C$ as a function of $N^S$ (a) and as function of $N^{AC} = N^S - N_f^C$ (b). The values of $N^S$ have been sampled from a Gaussian distribution centered at $<N^S> = 200$ and with standard deviation $\sqrt{\langle N^S \rangle} = \sqrt{200}$.

We have used the described experimental procedure to determine all the covariance terms needed to evaluate the uncertainty of $P_\gamma$. We obtain that the covariance terms involving $\varepsilon_\gamma$ are a factor $10^{-3}$ smaller than the other covariance terms and can also be neglected. Therefore, only one additional covariance term, $Cov(N_\gamma^{C,tot}; N_{\gamma,f}^C)$, has to be considered for the determination of the uncertainty on $P_\gamma$. The results of the covariance analysis are listed on Table 1.

| Covariance | $P_\gamma$ | $P_f$ |
|---|---|---|
| $Cov(N^S; N_\chi^C)$ | $\approx Var(N_\gamma^C)$ | $\approx Var(N_f^C)$ |
| $Cov(N^S; \varepsilon_\chi)$ | $\approx 0$ | $= 0$ |
| $Cov(N_\chi^C; \varepsilon_\chi)$ | $\approx 0$ | $= 0$ |
| $Cov(N_\gamma^{C,tot}; N_{\gamma,f}^C)$ | $\approx Var(N_{\gamma,f}^C)$ | - |
| $Cov(N_\gamma^{C,tot}; \varepsilon_f)$ | $= 0$ | - |
| $Cov(N_{\gamma,f}^C; \varepsilon_f)$ | $= 0$ | - |

**Table 1:** Covariance terms necessary to determine the uncertainty on the decay probabilities $P_\chi$, the index $\chi$ refers to either fission or gamma decay.

In our experiment, the probabilities were measured at different excitation energies with the same setup. This introduces a correlation between the probabilities measured at different energies, $E_i^*$ and $E_j^*$, which must be accounted for. As shown in [15]:

$$Corr\left(P_\chi(E_i^*); P_\chi(E_j^*)\right) = \sqrt{\frac{Var(P_\chi^{syst}(E_i^*)) \cdot Var(P_\chi^{syst}(E_j^*))}{Var(P_\chi(E_i^*)) \cdot Var(P_\chi(E_j^*))}} \quad \text{if} \quad i \neq j \qquad (9)$$
$$= 1 \quad \text{if} \quad i = j$$

where $Var(P_\chi^{syst}(E_i^*))$ corresponds to the systematic part of the total variance of the decay probability at energy $E_i^*$. Eq. (9) says that the correlation measures the importance of the systematic uncertainty with respect to the total uncertainty. It is close to 1 when the systematic uncertainty dominates the total uncertainty. In our measurement, the systematic uncertainty comes from the uncertainty on the fission detection efficiency and from the presence of contaminant peaks in the singles spectrum.

## IV Results and discussion

Fig. 10 shows the results for the gamma-decay and fission probabilities. As already mentioned, our setup allowed us to measure the decay probabilities at eight different angles. We observe a decrease of the gamma-decay probability with increasing angle, whereas for the fission probability we observe an increase in the region from 6.1 to 6.5 MeV. For the sake of clarity in Fig. 10 a) and b) we show only the decay probabilities measured at the limiting angles 126 and 140 degrees. We can see that the differences between the decay probabilities measured at these two angles are significant. The possible origin of these differences will be discussed in section IV B.

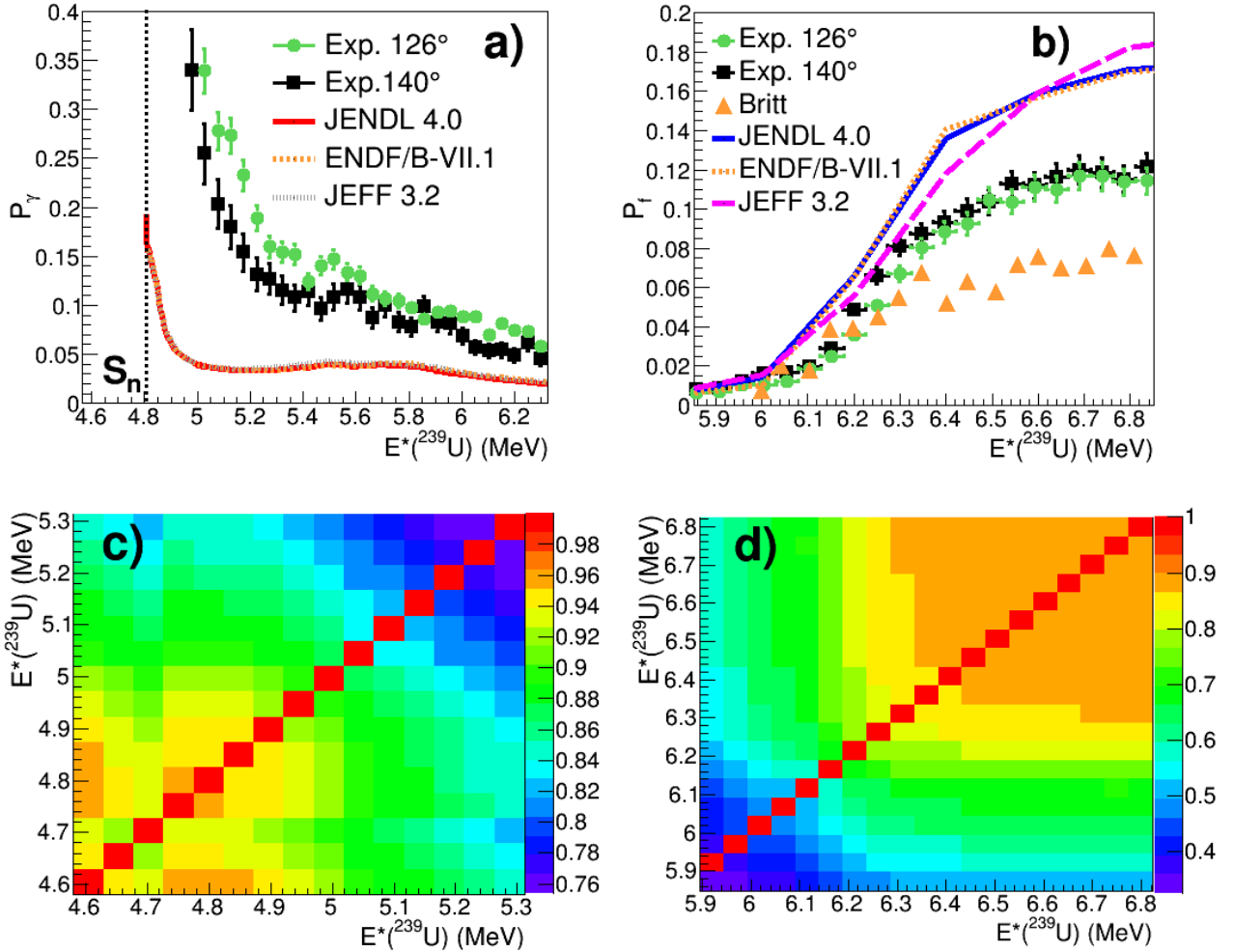

**Figure 10:** (Color online) Measured gamma-decay (a) and fission (b) probabilities as a function of excitation energy (symbols) compared to the results of several evaluations (lines). Correlation matrix for the gamma-decay (c) and fission (d) probabilities measured at 126 degrees. The vertical dotted line in panel a) represents the neutron separation energy of $^{239}U^*$.

In Fig. 10 panels c) and d) the correlation matrices for the gamma-decay and fission probabilities measured at 126 degrees are shown. They are representative of the correlation matrices for all the other detection angles. For the gamma-decay probability the correlation is the highest at the lowest excitation energies near $S_n$. This is due to the fact that the statistical uncertainty of the gamma-decay probability increases with excitation energy and the systematic uncertainty is larger near $S_n$, due to the presence of contaminant peaks in the singles spectrum. On the contrary, for the fission probability the correlation is the highest at high excitation energies. The reason is that the statistical uncertainty on the fission probability decreases with excitation energy and the systematic uncertainty on the fission efficiency gives the strongest contribution to the total uncertainty at the highest excitation energies.

In Figures 10 a) and b), our data are compared to the neutron-induced decay probabilities given by different evaluations. The latter have been obtained by dividing the evaluated neutron-induced cross sections by the compound-nucleus formation cross section $\sigma_{CN}$, according to eq. (1). $\sigma_{CN}$ was obtained with the phenomenological optical-model potential used in the JENDL 4.0 evaluation [19]. The gamma-decay probability obtained with the surrogate method is several times higher than the neutron-induced one over the whole excitation-energy range. The discrepancies between the surrogate data and the neutron-induced data decrease with excitation energy. A minimum factor of about 3 is reached near 6.3 MeV. The fission probability obtained with the surrogate reaction is in good agreement with the neutron-induced data below about 6 MeV. Above 6 MeV the JENDL and ENDF evaluations are in very good agreement and show significant differences with respect to the JEFF evaluation. Between 6 and 6.3 MeV our data are in better agreement with the JEFF evaluation. Above 6.3 MeV our results are systematically below the neutron-induced results. We observe differences up to 30-35%. The reason for the discrepancy with respect to the neutron-induced data may be the deuteron breakup, which leads to a background of "sterile" protons that contaminates the singles proton spectrum. These protons are not related to the formation of a compound nucleus $^{239}$U* and lead to a decrease of the measured fission probability, as shown by eq. (2). This hypothesis was already put forward by Britt and Cramer [7], but only now it starts to attract theoretical efforts [8, 9]. Interestingly, the data by Britt and Cramer [7] obtained using the same $^{238}$U(d,p) reaction with a beam energy of 18 MeV and protons detected at 150 degrees are 30% lower than our data at the fission plateau. The impact of deuteron breakup on the fission probability at 15 and 18 MeV incident energies is evaluated in section IV A.

Because the oxidation of the target could not be completely avoided, fusion of the deuteron beam with oxygen and the subsequent evaporation of protons have also to be taken into account. Again, this leads to the production of sterile protons in the excitation-energy range of interest, decreasing the measured fission probability. Therefore, this process might also be responsible for the differences observed between the surrogate data and the neutron-induced data, as well as between the two surrogate-reaction results. Indeed, as mentioned above, in our experiment we limited as much as possible the oxidation of the $^{238}$U metallic target, whereas the $^{238}$U target used by Britt and Cramer was an oxide. According to the PACE4 code [20], the kinetic energies of the protons originating from fusion-evaporation on oxygen correspond to equivalent excitation energies of $^{239}$U larger than 6.3 MeV. Using PACE4, we estimated that in our case the fraction of these protons is of the order of 10%. To obtain this value we used the number of oxygen atoms in the target that results from counting the number of elastically scattered deuterons on oxygen, which can be seen on Fig. 3, and using the corresponding Rutherford-scattering cross section. This estimation leads to a fraction of about two oxygen nuclei per three uranium nuclei in the target. Note that the contribution to the singles spectrum of the protons evaporated after the fusion of the deuteron with the carbon nuclei of the target backing is removed when the carbon-backing spectrum $N_c$ is subtracted from the proton spectrum $N_p$, see section III. B. We do not know the chemical composition of the target used by Britt and Cramer, but in view of the possible chemical forms of uranium oxide ($UO_2$, $UO_3$, $UO_4$ and $U_3O_8$) we can say that there were at least two atoms of oxygen per uranium atom. This corresponds to a factor 3 more oxygen than in our target and thus to at least about 30% of protons originating from fusion-evaporation reactions on oxygen. Therefore, the larger amount of oxygen in the target used by Britt and Cramer might explain, at least partly, the differences between the two sets of surrogate-reaction data.

**A Deuteron breakup**

In principle, the (d,p) reaction can be seen as a two-step process in which, first, the deuteron breaks up and then the neutron is absorbed by the target nucleus. However, this picture is way too simple. In fact, the deuteron breakup is a rather complex process. One has to distinguish between elastic and inelastic breakup. In the elastic breakup (EB), the impinging deuteron breaks up due to the Coulomb and/or nuclear interaction with the target, and the resulting proton and neutron move apart leaving the target nucleus in the ground state. The non-elastic breakup (NEB) includes the processes in which the incident deuteron breaks up and the resulting proton and neutron move apart leaving the target nucleus in an excited state; the direct stripping of the neutron, and the fusion of the breakup neutron with the target nucleus which leads to the compound-nucleus formation. The latter mechanism, that we will call breakup fusion (BF), is the one of interest in the context of the surrogate-reaction method. In an inclusive measurement, as ours, where only the proton is detected, it is not possible to experimentally discriminate the different processes. Therefore, here we rely on theory to estimate the contributions of the different mechanisms to the measured proton singles spectrum.

The breakup process was the subject of intense theoretical work in the eighties. Udagawa and Tamura [21] described the A(d,p) reaction within the distorted-wave Born approximation (DWBA) in prior form, whereas Ichimura, Austern and Vincent [22] used the post-form DWBA. The equivalence of the post and prior formulations was first demonstrated in the original work of Ichimura, Austern, and Vincent [22], although the prior-form formula derived in [22] differed from that proposed by Udagawa and Tamura. Both approaches have in common that the non-elastic breakup cross section is proportional to a matrix element $<\psi_n|W_{nA}|\psi_n>$ where $\psi_n$ is the wave function describing the evolution of the neutron and $W_{nA}$ is the imaginary part of the optical potential between the neutron and the nucleus $A$. In a recent publication, Potel et al. [8] discuss the equivalence between the post and prior methods and present results for the elastic and non-elastic breakup cross section of the $^{93}$Nb(d,p) reaction. However, this study does not give the separated contribution from BF.

In this work, the EB contribution has been obtained with the continuum-discretized coupled-channels (CDCC) method, using the coupled-channels code FRESCO [23]. We have used a model based on the method of Ichimura, Austern and Vincent to determine the NEB [9]. To estimate the BF part, the imaginary part of the potential $W_{nA}$ has been divided into two parts, a part $W_{nA}^{CN}$ corresponding to the compound-nucleus formation and a part associated to all the other remaining processes included in the NEB. $W_{nA}^{CN}$ was parametrized in terms of a Woods-Saxon form, with the parameters adjusted to reproduce the compound-nucleus formation cross section as predicted by the JENDL 4.0 evaluation [19]. The results of our calculations for 15 and 18 MeV deuteron incident energies and 140 and 150 degrees, respectively, are shown in Fig. 11. The results for 15 MeV and 126 degrees are not shown because they are very close to the results obtained for 140 degrees. The formalism we have used can be applied when the neutron ends up in a bound state (negative neutron energies) or in an unbound state (positive neutron energies). However, the calculations shown in Fig. 11 consider only transfer of the breakup neutron to unbound states. This is why we only show values of the cross sections for $E^* > S_n$.

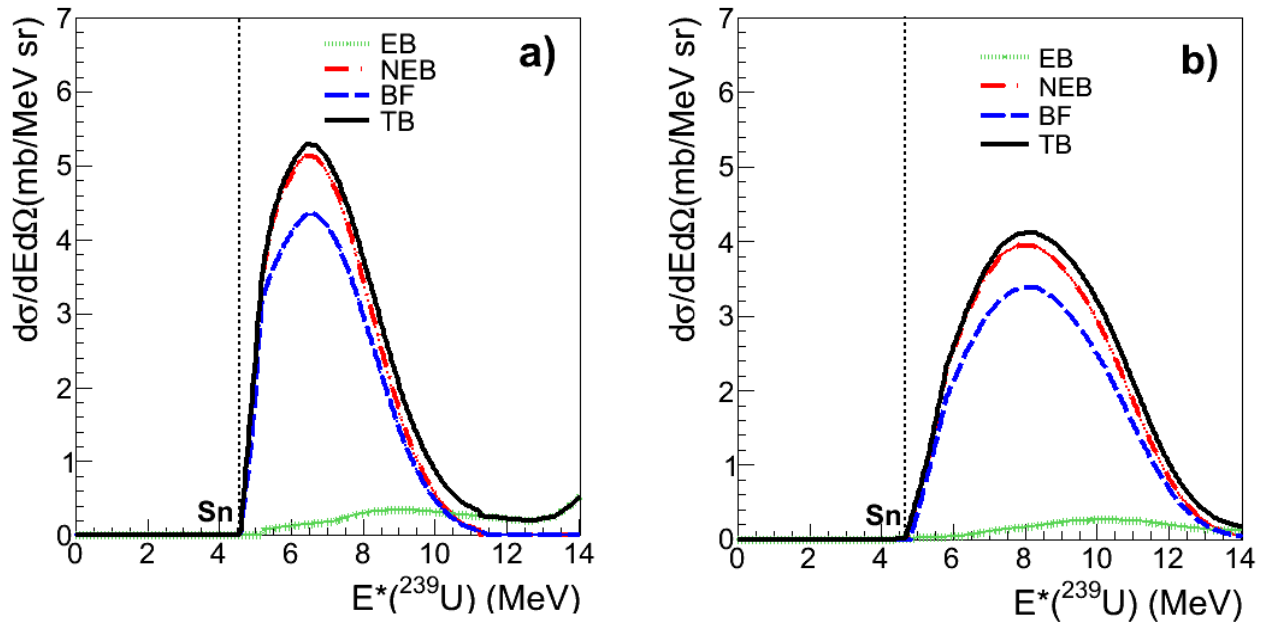

**Figure 11:** (Color online) Calculated contributions to the total deuteron breakup process (TB), as a function of the excitation energy of $^{239}$U for a deuteron beam energy of 15 MeV and a proton angle of 140 degrees (a) and for a beam energy of 18 MeV and proton angle of 150 degrees (b). NEB corresponds to non-elastic breakup, BF to breakup fusion and EB to elastic breakup. Note that TB = NEB + EB. The vertical dotted lines indicate the neutron separation energy of $^{239}$U.

It was argued in [24] that the approach of Udagawa and Tamura only gives the so-called "elastic breakup fusion", that is, the BF not accompanied by the simultaneous excitation of the target. Therefore, the Udagawa and Tamura approach gives a lower limit of the BF contribution, as it excludes processes in which a compound nucleus is formed after target excitation. This is indeed the case, since the BF cross section we obtained with the Udagawa and Tamura approach is about 4% smaller than the one obtained with the Ichimura, Austern and Vincent method used in ref. [9].

The relative contribution of the different processes to the total cross section for both incident energies and detection angles is rather similar, see Fig. 11. In the region of interest in this work, $E^* < (S_n +1.5)$ MeV, the elastic breakup represents less than 5 % of the total breakup, whereas the breakup fusion represents nearly 80 %. The total breakup (TB), given by the sum of the elastic and inelastic breakup, can be directly related to the proton singles spectrum above $S_n$. Therefore, these calculations allowed us to correct the singles spectrum from the sterile protons originating from elastic and non-elastic breakup. The corrected decay probabilities $P_\chi^{corr}$ have been obtained in the

following way:

$$P_\chi^{corr}(E^*) = \frac{P_\chi^{meas}(E^*) \cdot \sigma_{TB}(E^*)}{\sigma_{BF}(E^*)} \qquad (10)$$

where $P_\chi^{meas}$ is the measured decay probability. Eq. (10) implies the assumption that contributions other than BF to the NEB lead neither to fission nor to gamma emission. This is reasonable for the fission exit channel but it is rather probable that the $^{238}$U* and $^{239}$U* nuclei that are excited by the other processes emit gamma rays. Therefore, the corrected gamma-decay probability should be considered as an upper limit of the real gamma-decay probability.

The corrected decay probabilities measured at 140 degrees are presented in Fig. 12. Obviously, the disagreement between the gamma-decay probability obtained with the $^{238}$U(d,p) reaction and the neutron-induced data increases when the breakup correction is applied. On the other hand, the corrected average fission probability is in better agreement with the neutron-induced data. However, in the fission plateau, the corrected fission probability is still lower by about 15% than the neutron-induced data represented by the JENDL and ENDF evaluations. This difference may be attributed to the contribution from protons originating from fusion-evaporation on oxygen, described above. When the fission probability measured by Britt and Cramer [7] is corrected, the resulting fission probability is still significantly lower than the neutron-induced data. However, this does not mean that the breakup calculations for this case are incorrect, since the remaining difference might be attributed to protons coming from fusion-evaporation on oxygen, which can be numerous due to the complete oxidation of the target used in [7].

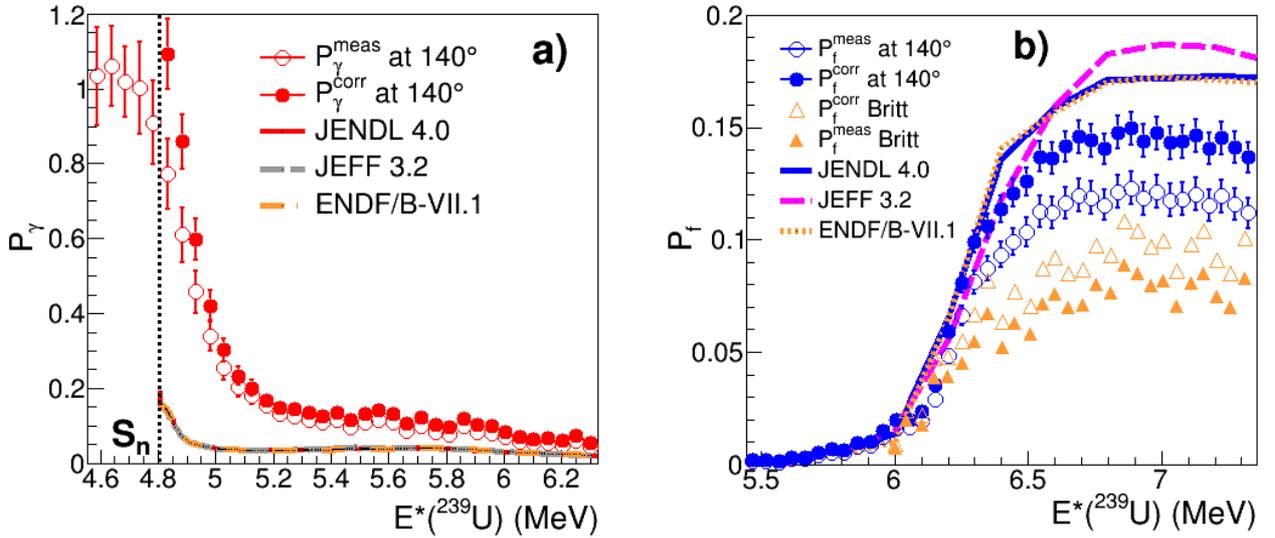

**Figure 12:** (Color online) Measured $P_\chi^{mes}$ and corrected $P_\chi^{corr}$ decay probabilities as a function of excitation energy. The neutron-induced decay probabilities from different evaluations are represented by the lines. The gamma-decay probabilities are shown in panel a) and the fission probabilities in panel b). The vertical dotted line in panel a) indicates the neutron separation energy of $^{239}$U.

## B Comparison with statistical model calculations

Figure 13 shows the breakup-corrected fission probability and the gamma-decay probability obtained in the surrogate reaction at 140 degrees together with the neutron-induced probabilities in the excitation-energy region where gamma emission and fission are in competition. In this energy range, the corrected fission probability is in good agreement with the neutron-induced data, whereas the gamma-decay probability obtained with the (d,p) reaction is several times higher than the neutron-induced one. We have chosen to present the uncorrected data for the gamma-emission probability in this figure to show that the discrepancies are not due to the breakup correction; they exist even for the lower limit of the gamma-emission probability. The objective of this section is to investigate whether we can explain this observation within the frame of the statistical model. For this purpose we have used the EVITA code, which is a Hauser-Feshbach Monte-Carlo code developed at the CEA DAM that uses the same ingredients as the TALYS code [25]. The parameters of the EVITA code are the ones of the JEFF evaluation shown in Figs. 10, 12 and 13, which were carefully tuned to reproduce the experimental neutron-induced data.

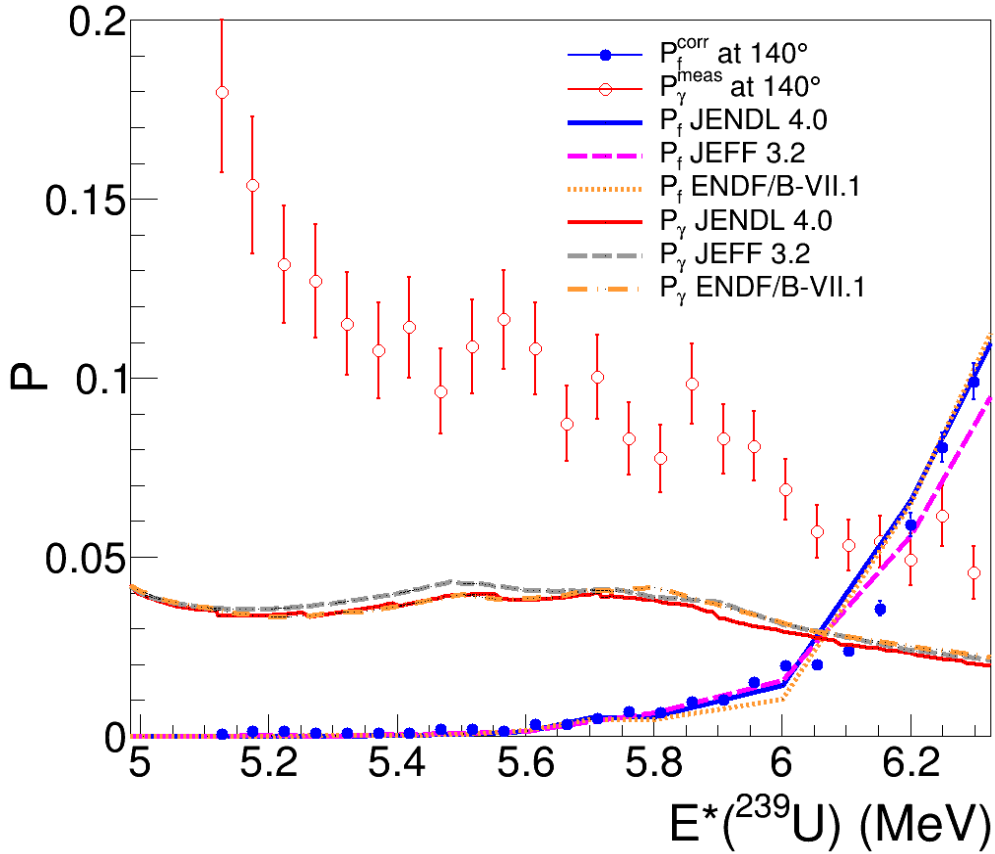

**Figure 13:** (Color online) Fission and gamma-decay probabilities as a function of excitation energy compared to the corresponding neutron-induced decay probabilities according to different evaluations. Note that only the fission probability has been corrected for deuteron breakup.

Figure 14 shows the "partial" gamma-decay and fission probabilities $P_\chi(E^*,J^\pi)$ calculated with the EVITA code for given initial values of spin and parity of the nucleus $^{239}U^*$ in the region where both decay channels compete. It is clear from Fig. 14 that the Weisskopf-Ewing approximation discussed in the introduction is not valid either for the gamma-emission or the fission probabilities. The calculated gamma-decay probabilities increase considerably with the spin of the decaying nucleus. This is mainly due to the hindering of neutron emission discussed in the introduction. On the other hand, the calculated fission probabilities decrease considerably with the angular momentum because the transition states on top of the fission barriers with the higher spins lie at higher excitation energies, leading to higher effective fission barriers. The calculations show that the parity of the states does not modify these major trends. However, for a fixed initial spin, a change in parity can have a significant impact on the associated decay probability for both fission and gamma emission.

Let us now go back to Fig. 10 where we compared the decay probabilities measured at 126 and 140 degrees and use the EVITA results from Fig. 14 for interpretation. One could explain the observed differences as the result of the variation of the populated spin distribution with the angle of the ejectile. Our results would suggest that the mean angular-momentum populated when the ejectile is emitted at 126 degrees is somewhat higher than the one populated when the ejectile is emitted at 140 degrees.

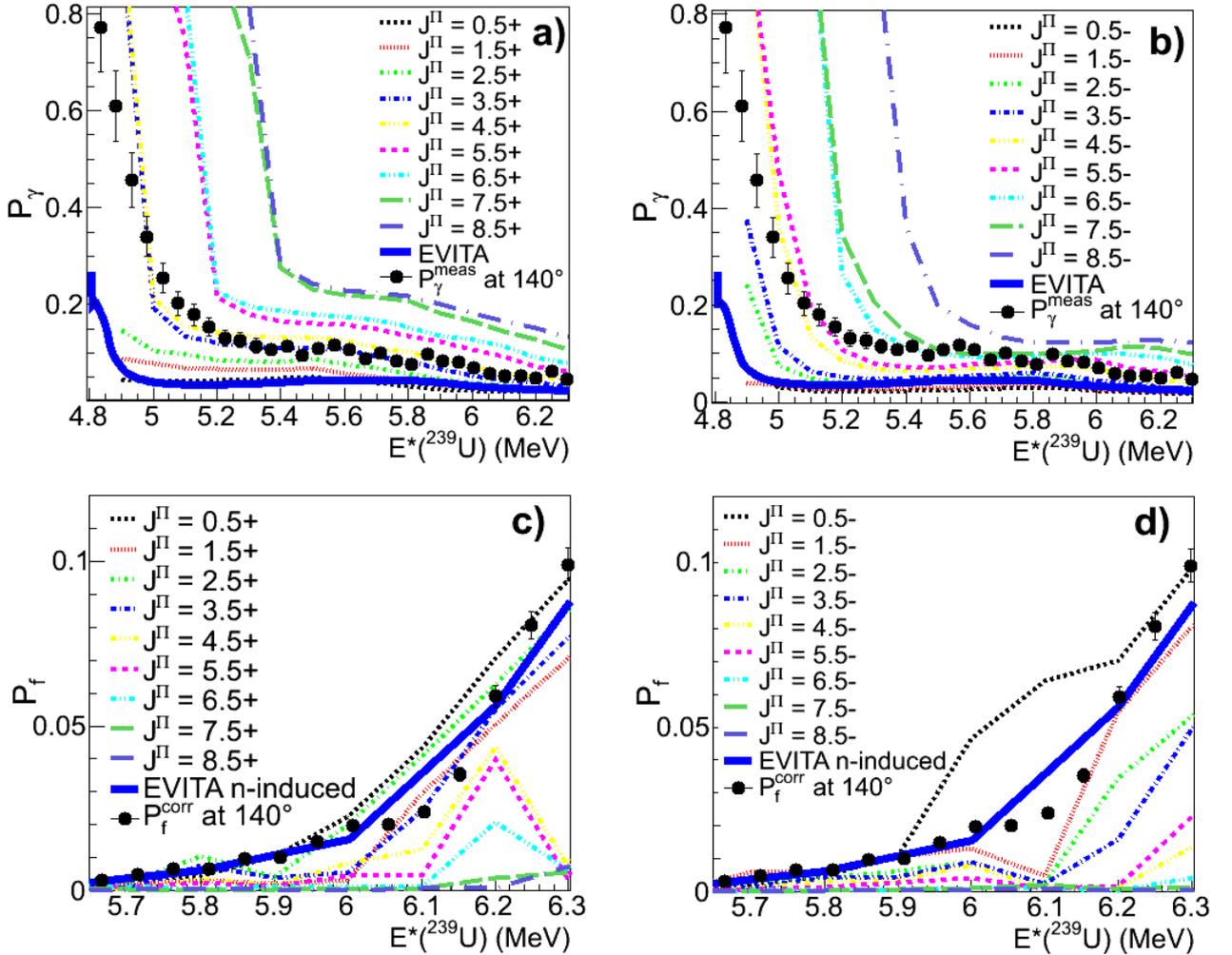

**Figure 14:** (Color online) Decay probabilities as a function of excitation energy for different values of angular momentum and parity of $^{239}$U* calculated with EVITA. The decay probabilities measured in this work (full circles) and the neutron-induced decay probabilities obtained with EVITA (thick blue lines) are also shown. Gamma-decay probabilities with positive and negative parities are shown in panels a) and b), respectively. Fission probabilities are shown in panels c) and d).

To calculate the decay probabilities it is necessary to weight the partial decay probabilities $P_\chi(E^*,J^\pi)$ by the probability to populate a given initial spin and parity, i.e. it is necessary to determine the initial spin and parity distribution populated in the $^{238}$U(d,p) reaction. Note that the spin-parity distribution that is relevant for the surrogate-reaction method is the one of the compound nucleus, see [2]. This means that, in the case of the (d,p) reaction, the spin-parity distribution of interest is the one associated to the breakup-fusion process. The spin distribution calculated by Potel et al. (ref. [8]) is not yet the relevant one because the model of Potel et al. does not separate the breakup-fusion component of the non-elastic breakup. We have used the model of [9] with the division of the optical potential into two parts (see section IV A) to obtain a first estimate of the average spin populated in the $^{238}$U(d,p) reaction for the breakup-fusion process. A complete calculation of the full shape of the spin and of the parity distribution with this formalism will be performed in the future. The results for the average spin $\bar{J}$ are shown in Table 2, where they are compared with the values for the distributions populated by the neutron-induced reaction n + $^{238}$U obtained with the optical-model potential used in the JENDL 4.0 evaluation [19]. We can see that the average spin populated in the (d,p) reaction is significantly larger than the one populated in the neutron-induced reaction, although the difference decreases with increasing excitation energy. At the lower excitation energies the average spin populated in the (d,p) reaction is about 71% larger than the one induced by neutron absorption, and it is about 23% larger at $E^* = S_n + 1.5$ MeV = 6.3 MeV. These calculations combined with the results of Fig. 13 indicate that the significant change in the spin distribution caused by the different entrance channel has a much stronger impact on the gamma-emission probability than on the fission probability. Interestingly, the calculations of Table 2 predict a slight increase of the average populated spin with decreasing angle, which is in line with the observed angular dependence of the decay probabilities shown in Fig. 10.

|  | $E^*=S_n+0.5$ MeV | $E^*=S_n+1$ MeV | $E^*=S_n+1.5$ MeV |
|---|---|---|---|
| $\bar{J}$ for $^{238}$U(d,p) at 140° | 2.4 | 2.5 | 2.6 |
| $\bar{J}$ for $^{238}$U(d,p) at 126° | 2.4 | 2.6 | 2.7 |
| $\bar{J} \pm \Delta J$ for n + $^{238}$U | 1.4±0.3 | 1.8±0.4 | 2.2±0.5 |

**Table 2:** Preliminary results for the average spin $\bar{J}$ populated in the $^{238}$U(d,p) reaction for the breakup-fusion component at different excitation energies and proton detection angles calculated with the model of [9]. The average spin and the RMS ($\Delta J$) populated in the n + $^{238}$U reaction obtained with the optical-model potential used in the JENDL 4.0 evaluation [19] is also given for comparison. $S_n$ is the neutron separation energy of $^{239}$U, which is 4.8 MeV.

In an attempt to obtain more information on the shape of the populated spin and parity distribution we have used the calculated partial decay probabilities $P_\chi(E^*,J^\pi)$ shown in Fig. 14 to fit the decay probabilities with the expression:

$$P_\chi(E^*) = \sum_{J^\pi} \left[ \frac{1}{2\sigma\sqrt{2\pi}} e^{-\frac{(J-\bar{J})^2}{2\sigma^2}} \right] \cdot P_\chi(E^*, J^\pi) \tag{11}$$

where the unknown angular-momentum distribution has been approximated with a Gaussian distribution without dependence on the excitation energy and the two parities are assumed to be equally populated. $\bar{J}$ and $\sigma$ correspond to the average value and the standard deviation of the spin distribution, respectively, and are free parameters. The values we obtained for these free parameters when we applied this fit procedure to the neutron-induced and the surrogate-reaction data are listed in Table 3. We can see that the values of $\bar{J}$ obtained from the fit to the neutron-induced decay probabilities calculated with EVITA are compatible with the calculated values from the optical potential of JENDL listed in the lower part of Table 2. The spin distributions for the $^{238}$U(d,p) reaction deduced from the fits to the measured fission and gamma-decay probabilities are clearly incompatible. The incompatibility is further demonstrated in Fig. 15 where the fission probability obtained with the spin distribution derived from the fit to the uncorrected gamma-decay probability and the $P_f(E^*,J^\pi)$ probabilities from EVITA is shown. This probability is clearly below our experimental data. The values of $\bar{J}$ obtained from the fit to the gamma-emission probability (Table 3) also differ considerably from the calculated values for the $^{238}$U(d,p) reaction (Table 2). These inconsistencies might be an indication that the equal population of positive and negative parities is not applicable to the $^{238}$U(d,p) reaction.

|  | $\bar{J}$ | $\sigma$ |
|---|---|---|
| $P_f^{EVITA,n}$ | 1.3 | 1.1 |
| $P_\gamma^{EVITA,n}$ | 1.5 | 1.2 |
| $P_f^{corr}$ 140° | 1.4 | 0.8 |
| $P_\gamma^{meas}$ 140° | 4.5 | 1.4 |
| $P_f^{corr}$ 126° | 2 | 1 |
| $P_\gamma^{meas}$ 126° | 5.4 | 1.2 |

**Table 3:** Values of the fit parameters obtained by fitting the decay probabilities with function (11).

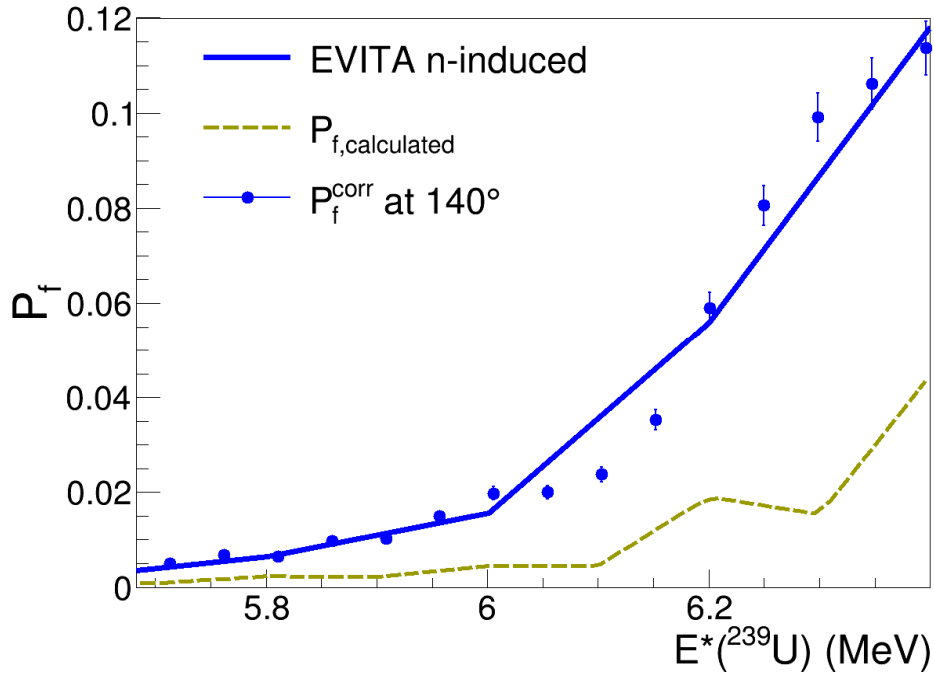

**Figure 15:** (Color online) Fission probability as a function of excitation energy. Our corrected data (dots) and EVITA results for the neutron-induced probability (full line) are compared to an EVITA calculation performed with the initial spin distribution deduced from a fit to our gamma-decay probability (dashed line).

# V Conclusion and perspectives

We have investigated the $^{238}$U(d,p) reaction by measuring, for the first time, the gamma-decay and fission probabilities simultaneously. Our fission probability is lower than the one deduced from the neutron-induced data. This difference is explained, to a great extent, by the contribution from elastic and inelastic deuteron breakup. Calculations of the elastic breakup following the continuum-discretized coupled-channels method and of the inelastic breakup obtained with the distorted-wave Born approximation have been used to correct our data. The corrected results agree with the neutron-induced data at the fission threshold but they are about 15% lower than the neutron-induced data at the fission plateau. This remaining difference can be explained by the contribution from protons evaporated after the fusion of the deuteron beam with the oxygen contamination in the target. Our gamma-decay probability is several times larger than the neutron-induced one. The discrepancy decreases as the excitation energy increases but it is still about a factor 3 at the highest excitation energies. The correction of the breakup contribution leads to even larger differences.

In the energy region where fission and gamma emission compete, the corrected fission probability measured for the $^{238}$U(d,p) reaction is in rather good agreement with the neutron-induced data, whereas the gamma-decay probability is several times higher than the neutron-induced one. We have used the Hauser-Feshbach code EVITA, which is based on TALYS and uses the parameters of the JEFF 3.2 evaluation, to interpret these results within the framework of the statistical model. Our statistical-model calculations predict a strong sensitivity of the gamma-emission and fission probabilities to the angular momentum. This implies that the Weisskopf-Ewing approximation is not applicable either to gamma-emission or to fission in the considered excitation-energy range. The model of ref. [9] modified to account for the breakup-fusion process has been used to obtain a first estimate of the average spin populated by the $^{238}$U(d,p) reaction. The latter average spin is between 71 to 23% larger than the average spin induced in the n+$^{238}$U reaction. We therefore conclude that none of the two limiting situations described in the introduction can explain our results.

The present work indicates that the fission probability is much less sensitive to the populated angular momentum than the gamma-decay probability. In the future, we will investigate whether we can explain this with our Hauser-Feshbach calculations by using the initial spin *and* parity distribution populated by the $^{238}$U(d,p) reaction that will result from the model of [9]. Unfortunately, the deuteron breakup complicates significantly the interpretation of our results. For this reason we have performed a measurement with a $^3$He beam on $^{238}$U to investigate the transfer reactions $^{238}$U($^3$He,t) and $^{238}$U($^3$He,$^4$He) which do not suffer from the breakup process. The simultaneous determination of the fission and gamma-decay probabilities for these reactions according to the method developed in this work shall provide a stringent test of the ingredients of the statistical model and considerably help in the understanding of the surrogate-reaction method.

# Acknowledgments

We would like to thank J. C. Müller, E. A. Olsen, A. Semchenkov, and J. C. Wikne from the Oslo Cyclotron Laboratory for providing the deuteron beam during the experiment and the GSI Target Laboratory for the production of the $^{238}$U target. This work was supported by the European Commission within the 7$^{th}$ Framework Program through Fission-2010-ERINDA (Project No.

269499). A.C.L. acknowledges support from the Research Council of Norway, project grant no. 205528, and from the ERC-StG-2014, under grant agreement no. 637686. F.G. and S.S. gratefully acknowledge support from the Research Council of Norway, grant no. 20007. J.L. was partially supported by a grant funded by the China Scholarship Council. M.W. acknowledges support from the National Research Foundation of South Africa under grant no. 83867.